\documentclass[aps,prb,twocolumn,superscriptaddress,amsmath,floatfix,11pt,reprint]{revtex4-2} 
\usepackage[utf8]{inputenc}
\usepackage[T1]{fontenc}
\usepackage{graphicx}  
\usepackage{dcolumn}   
\usepackage{physics}
\usepackage{bm}        
\usepackage{amssymb}   
\usepackage{framed}
\usepackage{amsmath}
\usepackage{multirow}
\usepackage{hhline}
\usepackage{color}
\usepackage{xcolor}
\definecolor{bluegreen}{rgb}{0,0.2,0.8}
\usepackage{subfigure,amsmath,verbatim,moreverb}
\usepackage{tabularx}
\usepackage{adjustbox}
\usepackage{lipsum}
\usepackage{longtable}
\usepackage{booktabs}
\usepackage{adjustbox}
\usepackage{hyperref}
\UseRawInputEncoding
\usepackage{graphicx}
\usepackage{booktabs}
\usepackage{siunitx}
\usepackage[table]{xcolor}
\definecolor{lightcyan}{rgb}{0.88,1,1}

\usepackage[normalem]{ulem}  

\usepackage{etoolbox}
\AtBeginEnvironment{align}{\setcounter{subeqn}{0}}
\newcounter{subeqn} %

\begin{document}

\title{Physically motivated iso-orbital indicator for meta-GGA exchange functionals}
\author{Jeet Sharma}
\email{jeet.sharma@niser.ac.in}
\affiliation{School of Physical Sciences, National Institute of Science Education and Research, An OCC of Homi Bhabha National Institute, Jatni 752050, India}
\author{Abhishek Bhattacharjee}
\affiliation{School of Physical Sciences, National Institute of Science Education and Research, An OCC of Homi Bhabha National Institute, Jatni 752050, India}
\author{Bikash Patra}
\affiliation{Department of Physical and Applied Sciences
, Indian Institute of Information Technology Surat, Kholvad campus, Kamrej 394190, Surat (Gujarat), India}

 \author{Prasanjit Samal}
 \email{psamal@niser.ac.in}
 \affiliation{School of Physical Sciences, National Institute of Science Education and Research, An OCC of Homi Bhabha National Institute, Jatni 752050, India}

\begin{abstract}
The iso-orbital indicator $\alpha = (\tau - \tau^\mathrm{vW})/\tau^\mathrm{UEG}$ is a key ingredient of meta-generalized gradient approximation (meta-GGA) functionals, but diverges in low-density tails
, causing unphysical exchange potentials and systematic band gap errors as noted in [J. Chem. Phys. 150, 161101 (2019)]. We replace the denominator of $\alpha$ with a physically motivated Pauli KED drawn from the orbital-free DFT literature, eliminating the divergence in the low density atomic tail without any empirical regularization parameter. Testing two such enhancement factors: LKT and PGS, within the r$^2$SCAN and MS2 exchange functionals, we find that the modified indicators suppress spurious oscillations in the semilocal exchange potential and restore correct electron localization in atomic tails. For a ten-member cubic semiconductor benchmark, the band gap mean absolute error is reduced by 41.1\% for r$^2$SCAN@PGS and 48.8\% for MS2@PGS, while cohesive energy accuracy is largely preserved. The consistent improvement across two functionals with distinct constructions confirms a physical rather than functional specific origin, and motivates further development of meta-GGA functionals with constraint satisfying iso-orbital indicators.

\end{abstract}

\maketitle

Density functional theory (DFT) is, in principle, an exact formalism 
for the ground state energy and electron density of an interacting 
system of electrons under a scalar external potential. It is 
conventionally implemented through the Kohn-Sham (KS) scheme, in 
which an interacting many electron problem is mapped onto a fictitious 
non interacting system of electrons moving in an effective 
potential~\cite{KS-equations,Parr1989,PerdewKurth2003}.
In this mapping, all many body effects are incorporated into the exchange correlation (XC) energy functional, which must be approximated in practical calculations. Among the wide variety of XC approximations, the local density approximation (LDA) \cite{PerdewWang1992,SunPerdewSeidl2010,LSD_Perdew_Zunger_PhysRevB.23.5048}, generalized-gradient approximations (GGAs), most widely used
Perdew-Burke-Ernzerhof (PBE) functional \cite{PerdewBurkeErnzerhof1996} offer a favorable accuracy-cost tradeoff, but the highest semilocal rung, the meta-GGA (MGGA), achieves significantly improved accuracy 
by additionally incorporating the KS kinetic energy density (KED) $\tau$~\cite{Sun2015SCAN,Tao2016}. The additional ingredient $\tau$  encodes local orbital-overlap information absent in LDA and GGAs, 
and enables MGGAs to simultaneously describe molecules, solids, and surfaces~\cite{Sun2015SCAN,Furness2020r2SCAN}. Its inclusion arises naturally from the Taylor expansion of the exact spherically averaged exchange hole~\cite{Becke1983,Becke1993} and allows construction of correlation functionals that are exactly self-interaction free for any one-electron system~\cite{Becke1993}. The correct asymptotic behavior of the exchange potential in finite systems also relies critically on the quality of $\tau$ in the low-density tail~\cite{Jana_2018_KED_JCP,Bikashda_Pauli-KED_2019_PRB}.

One of the efficient ways the $\tau$ is introduced in MGGA is via the iso-orbital indicator, which characterizes the extent of local orbital overlap~\cite{Sun2012,Becke1983}. 
Early examples include $z = \tau^\mathrm{vW}/\tau$, which correctly identifies single-orbital ($z=1$) and slowly varying ($z\to 0$) regions but cannot distinguish the latter from noncovalent closed-shell overlap~\cite{Sun2013,Szymon_2019_computation7040065}, and $t^{-1} = \tau/\tau^\mathrm{UEG}$, which differentiates covalent and noncovalent bonding but does not uniquely identify single-orbital regions, contributing to numerical instability in empirical MGGAs such as M06-L~\cite{Sun2013,Zhao2006M06L}. The physically grounded and widely adopted indicator~\cite{Sun2015SCAN} is
\begin{equation}
    \alpha = \frac{\tau - \tau^\mathrm{vW}}{\tau^\mathrm{UEG}},
    \label{eq:alpha}
\end{equation}
where $\tau^\mathrm{vW} = |\nabla\rho|^2/(8\rho)$ is the von Weizs\"{a}cker KED~\cite{vonWeizsacker1935} and $\tau^\mathrm{UEG} = (3/10)(3\pi^2)^{2/3}\rho^{5/3}$ is the uniform electron gas KED. The indicator $\alpha$ clearly identifies single-orbital regions ($\alpha\approx 0$), slowly varying regions ($\alpha\approx 1$), and noncovalent overlap regions ($\alpha\to\infty$), and plays a central role in nonempirical MGGAs such as SCAN~\cite{Sun2015SCAN,Perdew2014Gedanken}. \\
%


\section{Numerical challenges and existing regularizations \label{sec:Challenges}}

Despite the success of modern MGGAs~\cite{Zhang2018Stability,Sun2016StructureEnergy,
Furness2018Cuprates,Yang2016Bandgap}, numerical instabilities in self-consistent 
field (SCF) calculations are well documented~\cite{Johnson2009Oscillations,
Yao2017grid,Szymon_2020_PhysRevB.101.165144,review_KED_Della_Sala_2016}, arising from rapidly varying derivatives of $\alpha$ in low-density regions. Bartok and Yates showed that these oscillations cause instabilities in pseudopotential generation and Fourier grid 
representations, making consistent plane wave calculations 
impractical~\cite{Bartok2019rSCAN}.

Furness and Sun~\cite{MS2_beta_Furness2019_isoorbital} proposed an alternative indicator
\begin{equation}
    \beta = \frac{\tau - \tau^\mathrm{vW}}{\tau + \tau^\mathrm{UEG}},
    \label{eq:beta}
\end{equation}
which has significantly smoother derivatives and eliminates the 
tail divergence. However, $\beta$ compresses the chemical environment 
identification that makes $\alpha$ powerful, the slowly varying limit 
shifts from $\alpha\approx 1$ to $\beta\approx 1/2$, and the 
noncovalent limit changes from $\alpha\to\infty$ to $\beta\to 1$ 
(saturation), while only the single orbital limit $\alpha=\beta=0$ 
is preserved (Table~\ref{alpha_comparison}). Consequently, any functional 
built on $\beta$ must retune its interpolation functions. Furness 
and Sun themselves noted that MS2$\beta$ was intended only as 
\textit{``a convenient means for preliminary investigation rather than 
a viable general functional''}~\cite{MS2_beta_Furness2019_isoorbital}.

The r$^2$SCAN functional~\cite{Furness2020r2SCAN} instead introduced a 
regularized $\alpha$,
\begin{equation}
    \bar{\alpha} = \frac{\tau - \tau^\mathrm{vW}}
    {\tau^\mathrm{UEG} + \eta\,\tau^\mathrm{vW}},
    \quad \eta = 10^{-3},
    \label{eq:alpha_bar}
\end{equation}
which preserves the physical range $0\leq\bar{\alpha}<\infty$ and the 
$\alpha\approx 0,~1,~\infty$ identification without retuning. However, 
$\eta$ was chosen purely for numerical  stability~\cite{Furness2020r2SCAN}, leaving room for improvement through physically motivated choices. 

We retain the exact orbital-dependent Pauli KED $\tau - \tau^\mathrm{vW}$ 
in the numerator and replace only the denominator with established and well tested 
GGA-level KED from the orbital free DFT (OFDFT)
literature~\cite{LKT,PGSL_2018_doi:10.1021/acs.jpclett.8b01926}, constructing denominators of the form  $\tau^\mathrm{UEG}F_\theta(s) + \tau^\mathrm{vW}$. This targets the  specific deficiency causing tail-region divergence without modifying the orbital-dependence or the potential calculation.


At this point, we emphasize that the importance of KED in exchange potential has been demonstrated by the work of Tran and Blaha~\cite{Tran2009TBmBJ} and further by Patra et.al. ~\cite{Jana_2018_KED_JCP,Bikashda_Pauli-KED_2019_PRB} within the meta-GGA framework. 
Their subsequent large scale benchmark~\cite{Tran_BG_benchmark_solids} confirmed that this
KED dependent construction is the only semilocal route to hybrid-functional accuracy for band gaps in solids, identifying the iso-orbital indicator as the lever controlling band gap quality at the semilocal level. Unlike potential-only constructions such as the modified Becke Johnson potential~\cite{Tran2009TBmBJ}, which sacrifice access to total energies, our modification operates entirely within the energy functional framework, providing band gap improvement alongside consistent cohesive energies and structural properties. This confirms that tail-region divergence of $\alpha$ has direct consequences for the exchange potential.

\begin{figure}
    \centering
    \includegraphics[width=.95\linewidth]{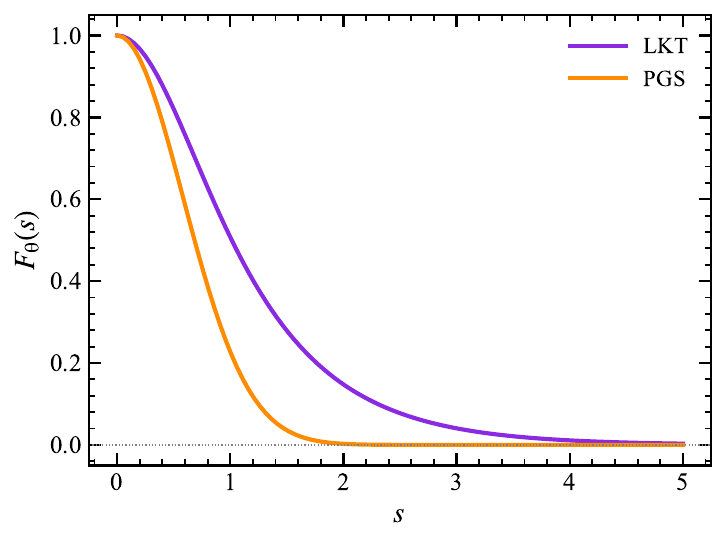}
    \caption{Pauli kinetic energy enhancement factors $F_{\theta}^{\mathrm{LKT}}$ and
             $F_{\theta}^{\mathrm{PGS}}$ as functions of the reduced density gradient $s$.
             Both satisfy $F_{\theta}(0)=1$ and decay monotonically to zero as
             $s\to\infty$, ensuring the modified iso-orbital indicator remains bounded
             in the atomic tail where $s\to\infty$.}
    \label{fig:F_theta_enhance}
\end{figure}

\section{Physical motivation and scope \label{sec:Motivation}}

The numerator of any iso-orbital indicator, $\tau-\tau^{\mathrm{vW}}$,
represents the Pauli KED $\tau_\theta$, the portion of the kinetic
energy arising from the Pauli exclusion principle beyond the
single orbital contribution~\cite{Levy1988PauliPotential}. The
denominator provides a reference scale normalizing this Pauli
contribution. In the standard $\alpha$ (Eq.~\ref{eq:alpha}), this
reference is $\tau^{\mathrm{UEG}}$, exact for the uniform electron gas
but with no mechanism to account for density inhomogeneity. In
$\bar{\alpha}$ (Eq.~\ref{eq:alpha_bar}), the $\eta\tau^{\mathrm{vW}}$
floor provides numerical regularization without physical motivation.

Thus, we here explore better reference energies to scale the Pauli factor while keeping the regularization intact. The Pauli KED is well studied in the OFDFT
literature~\cite{review_article_OFDFT_LargeScale,LKT}, where GGA-level enhancement
factors $F_\theta(s)$ have been constructed to satisfy known
constraints on $T_\theta$: non-negativity of the Pauli energy
($T_\theta[\rho]\geq 0$), non-negativity of the Pauli potential
($v_\theta(\mathbf{r})\geq 0$), and correct asymptotic
behavior~\cite{LKT,PGSL_2018_doi:10.1021/acs.jpclett.8b01926}.
In semilocal OFDFT, the total kinetic energy takes the form
$T_s = T_{\mathrm{vW}} + T_\theta$~\cite{review_article_OFDFT_LargeScale}, where the
Pauli contribution is
\begin{equation}
    T_\theta = \int \tau^{\mathrm{UEG}}(\mathbf{r})\,
    F_\theta(s(\mathbf{r}))\,\mathrm{d}^3r,
\end{equation}
with $F_\theta(s)$ the dimensionless Pauli enhancement factor and
$s = |\nabla\rho|/[2(3\pi^2)^{1/3}\rho^{4/3}]$ the reduced density
gradient. We explore two best~\cite{review_article_OFDFT_LargeScale} such candidate enhancement factors
(Fig.~\ref{fig:F_theta_enhance}):
\begin{align}
    F_\theta^{\mathrm{LKT}} &= \frac{1}{\cosh(as)},
    \quad a = 1.3~\cite{LKT}, \\
    F_\theta^{\mathrm{PGS}} &= \exp(-\mu s^2),
    \quad \mu = 40/27~\cite{PGSL_2018_doi:10.1021/acs.jpclett.8b01926}.
    \label{eq:Ftheta_LKT_PGS}
\end{align}
Both satisfy $0 \leq F_\theta \leq 1$ (Lieb bound~\cite{Lieb1980,Gazquez1982}) and recover $F_\theta \to 1$ as $s \to 0$, ensuring the modified indicator reduces to the standard $\alpha$ in the uniform density limit. Among the many enhancement factors available in the OFDFT literature, LKT and PGS represent two physically distinct constructions: LKT is built to guarantee
non-negativity of the Pauli potential, while PGS recovers the exact second order gradient expansion at small $s$. The two factors differ in their rate of decay: $F_\theta^{\mathrm{PGS}}$ reaches near zero around $s \approx 1$, corresponding to the onset of the strongly inhomogeneous regime, while $F_\theta^{\mathrm{LKT}}$ decays more slowly. As shown in Fig.~\ref{fig:F_theta_enhance}, this difference in decay rate
directly determines which parent functional benefits more from each factor and explains the contrasting impact on band gaps and structural properties.

\begin{table}[h]
\centering
\caption{Limiting values of iso-orbital indicators in physically 
important regions. The noncovalent bonding and slowly-varying 
values for $z$, $\alpha$, $\bar{\alpha}$, and $\beta$ are from 
Ref.~\cite{MS2_beta_Furness2019_isoorbital}; asymptotic 
expressions for the HOMO tail rows are derived in 
Appendix~\ref{app:asymptotic} using the exact KS KED 
decomposition~\cite{DellaSala2015KED}.}
\label{alpha_comparison}
\renewcommand{\arraystretch}{1.5}
\setlength{\tabcolsep}{6pt}
\begin{tabular}{lccccc}
\hline\hline
Region 
  & $z$
  & $\alpha$ 
  & $\bar{\alpha}$ 
  & $\beta$ 
  & $\alpha_{r^2@\mathrm{OF}}$ \\
\hline
Single orbital   
  & 1 & $0$ & $0$ & $0$ & $0$ \\
Slowly varying   
   & 0 & $1$ & $1$ & $\tfrac{1}{2}$ & $1^{\rm a}$ \\
Noncovalent bonding  
   & 0 & $\infty$ & $\infty^{\rm b}$ 
   & $\in (\frac{1}{2}, 1)$  & $\infty^{\rm e}$ \\
Tail ($s$-HOMO)  
   & 0 & $0$ & $0$ & $0$ & $0$ \\
Tail ($p$-HOMO)  
   & 0 & $\infty$ & $\gg 1^{\rm c}$ 
  & $0^{\rm d}$ & $0^{\rm d}$ \\
\hline\hline
\end{tabular}
\begin{flushleft}
{\footnotesize
$^{\rm a}$~Follows from $F_\theta(0)=1$~\cite{LKT,Karasiev2014VT84F};
unlike $\beta$, no reparametrization needed.\\
$^{\rm b}$~In noncovalent regions $\tau^{\mathrm{vW}}\to 0$, so the 
$\eta\tau^{\mathrm{vW}}$ floor vanishes and $\bar{\alpha}$ diverges 
like $\alpha$~\cite{Furness2020r2SCAN}.\\
$^{\rm c}$~Formally $\bar{\alpha}\to 0$ but reaches 
$\mathcal{O}(10^3)$ at intermediate $r$ for $\eta=10^{-3}$; 
see Eq.~(\ref{eq:app:alphabar-large}).\\
$^{\rm d}$~Both decay as $2/(\kappa^2 r^2)$; see 
Eqs.~(\ref{eq:app:beta-ptail}),~(\ref{eq:app:alphaOF-ptail}).\\
$^{\rm e}$ diverges only at the strict $\tau^{\mathrm{vW}} = 0$ bond center,
where $s\xrightarrow[]{}0$ forcing $F_\theta\xrightarrow[]{}1$~(Appendix A.4) .
}
\end{flushleft}
\end{table}

In the atomic tail, $s \to \infty$ and $F_\theta \to 0$ by construction, so the denominator of Eq.~\ref{eq:iso-orb-OF} is dominated by $\tau^{\mathrm{vW}}$ and remains finite even as
$\tau^{\mathrm{UEG}} \sim \rho^{5/3} \to 0$. The divergence of the standard $\alpha \sim \rho^{-2/3}$ is thereby eliminated without any empirical parameter, the cure is a consequence of the algebraic decay of $F_\theta$, not of its kinetic energy accuracy in the tail.
This motivates the proposed indicator

\begin{equation}
    \alpha_{\mathrm{\mathrm{r2@OF}}} = \frac{\tau - \tau^{\mathrm{vW}}}
    {\tau^{\mathrm{UEG}} F_\theta^{\mathrm{OF}} + \tau^{\mathrm{vW}}},
    \label{eq:iso-orb-OF}
\end{equation}

where r2@OF denotes replacement of the r$^2$SCAN  denominator with an
OF-KEDF form. The single-orbital limit ($\alpha_{\mathrm{r2@OF}} = 0$)
and slowly varying limit ($\alpha_{\mathrm{r2@OF}} \approx 1$, since
$F_\theta(0) = 1$) are both preserved without any reparametrization
of the interpolation functions, which is the key advantage over $\beta$
(Table~\ref{alpha_comparison}).


\section{Computational Details}
All the DFT calculations have been conducted using the VASP software package~\cite{Kresse1993} using plane wave basis and PAW pseudopotentials ~\cite{Blochl1994PAW,Kresse1999PAW}. For a complete test set on semiconductors, we use Si and nine III-V semiconductors (AlAs, AlP, AlSb, GaAs, GaP, GaSb, InSb, InP, InAs) in Fig.~\ref{fig:SC_R2S} and Fig.~\ref{fig:boxplot_MS2}. In addition, we have calculated cohesive energies on LC23 ~\cite{OFR2} data set and presented a comprehensive analysis in Table.~\ref{tab:cohesive} and Fig.~\ref{fig:cohesive}.

For the assessment of the structural properties, we calculate equilibrium volume (V$_0$) and bulk modulus (B$_0$) from energy vs volume curves by varying the experimental lattice constant $-10\%$ to $+10\%$. The resulting energy volume data obtained after convergence were fitted to the Birch-Murnaghan equation of state of third order using \textsc{VASPKIT}~\cite{WANG2021108033}. The plane wave cutoff energy (ENCUT) was set to 520~eV along with energy difference criterion in the SCF cycle to $10^{-6}$~eV for  convergence of all systems, and a $10 \times 10 \times 10$ Monkhorst-Pack k-point mesh was used. 

Cohesive energy was calculated by the standard procedure ~\cite{Kresse1993} of placing the atoms in a large simulation box of volume $17 \times 17 \times 17$ \AA $^{3}$ of vacuum to isolate from adjacent periodic blocks using Gamma k-point mesh of $1 \times 1 \times 1$. Next bulk calculation was  carried out, to evaluate energies at experimental lattice constant with Gamma k-point mesh of $10 \times 10 \times 10$. for consistency, ENCUT was set to be 520 eV and convergence was achieved by setting energy difference in SCF cycle to be $10^{-6}$ eV.

For the calculation of the band gaps of semiconductors, the experimental lattice constants were used with ENCUT set to 520~eV and an $10 \times 10 \times 10$ Gamma k-point mesh. Convergence was achieved by setting the energy difference criterion in the SCF cycle to $10^{-6}$~eV.
For the radial calculation of exchange potential, Hartree-Fock orbitals were used and the atomic orbitals were obtained by the most accurate data set provided by Koga and coworkers~\cite{KOGA1999}.
The computation for Li and C were executed with spin polarization using PySCF~\cite{pyscf_2020}, with the basis set def2-QZVP~\cite{weigend2005balanced}. The quantities $\rho$,$\nabla\rho$,$\tau$ were computed on grid in radial direction with 6000 points and subsequently used to construct the iso-orbital indicators and the corresponding ELFs.

\section{Results AND DISCUSSIONS \label{sec:Results} }

\begin{figure*}[htp]
    \centering
    
    \subfigure{
        \includegraphics[width=1.0\linewidth]{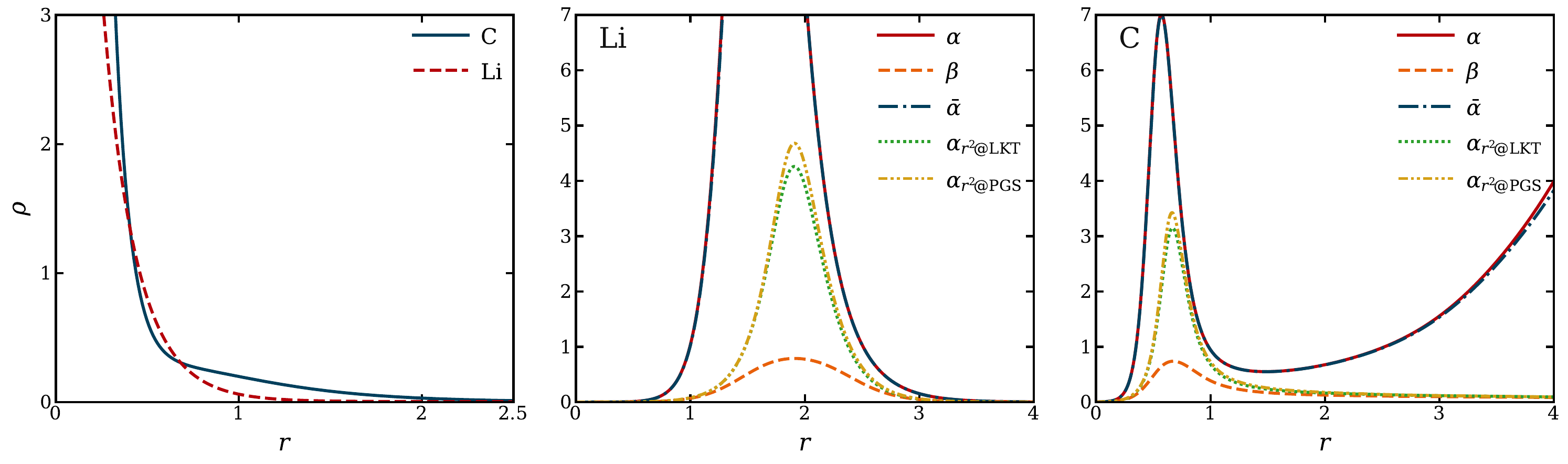}
    }\quad

    \caption{Comparison of electron densities and iso-orbital indicators for Li and C atoms.  
     Spherically averaged electron densities (a)$\rho(r)$ of Li and C atoms as
             functions of radial distance $r$ (in Bohr).  The extended density of C, reflects its larger nuclear charge, produces a richer shell structure and a more structured $s(r)$ profile, leading to qualitatively different behavior of the iso-orbital indicators in the tail. Iso-orbital indicators as functions of radial distance $r$ for
             (b) Li and (c) C atoms. }
    \label{fig:iso}
\end{figure*}

\begin{figure*}[htp]
    \centering
    \subfigure[Li atom: Electron localization function $1/(1+\alpha^{2})$]{
        \includegraphics[width=.40\linewidth]{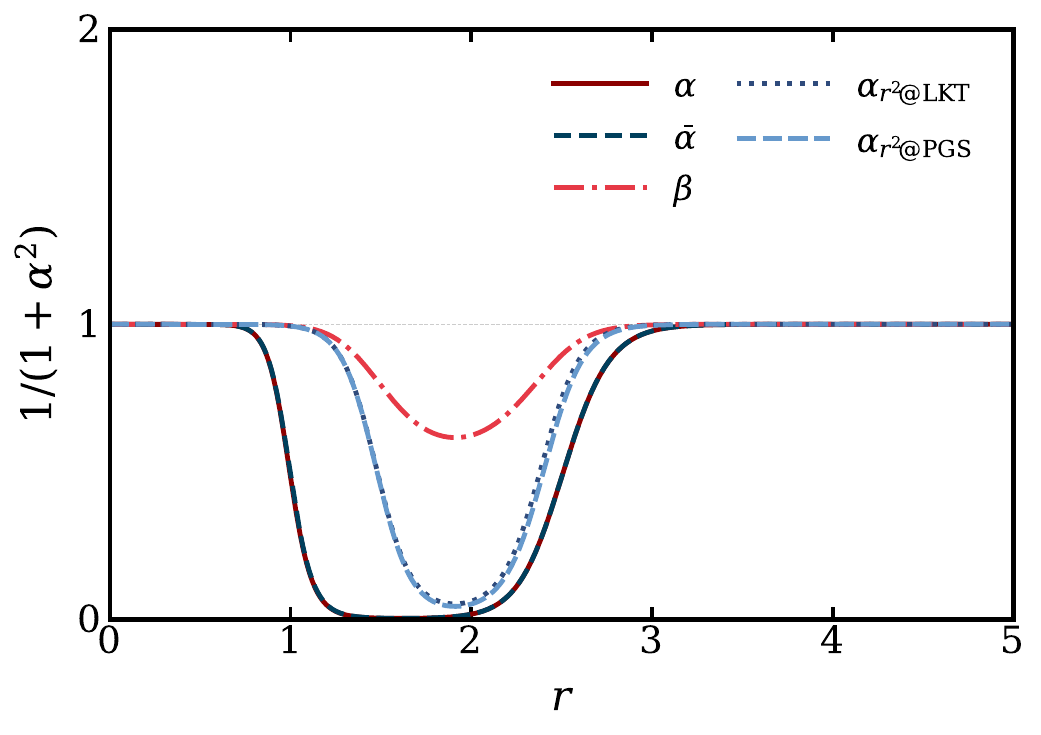}
    }\quad
    \subfigure[C atom: Electron localization function $1/(1+\alpha^{2})$]{
        \includegraphics[width=.40\linewidth]{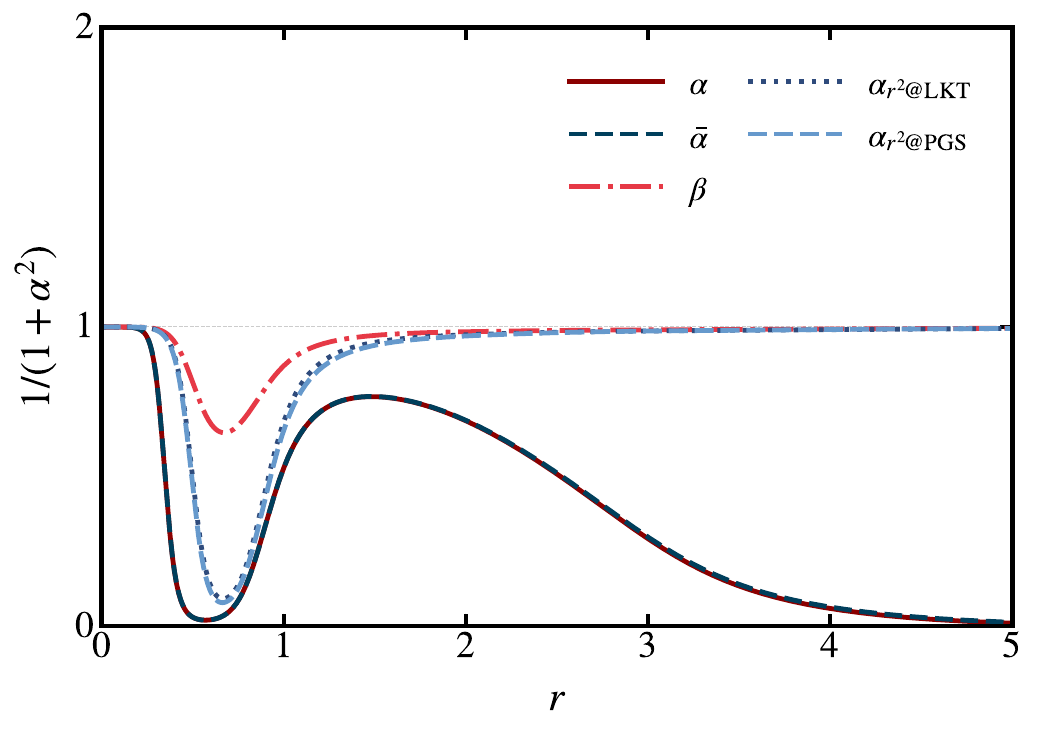}
    }
    \caption{Comparison of the electron localization function $1/(1+\alpha^{2})$ obtained from various iso-orbital indicators for spherically symmetric atoms. The curves illustrate how different indicators behave across the radial coordinate $r$, highlighting differences in the core, valence, and tail regions. For Li (s-type HOMO), 
all indicators recover the correct limit. For C (degenerate p-type HOMO), 
$\alpha$ and $\bar{\alpha}$ fail and give $f_{\mathrm{ELF}} \to 0$, 
while $\alpha_{r^2@\mathrm{OF}}$ and $\beta$ correctly restore 
$f_{\mathrm{ELF}} \to 1$.} 
    \label{fig:ELF}
\end{figure*}


\subsection{ The tail-tale of ELF \label{sec:Results:elf-iso-orb}}

In this section, we focus on the quality of the modified iso-orbital
indicators as descriptors of electronic structure. We compare the proposed indicators against $\alpha$, $\bar{\alpha}$, and $\beta$ across atoms, examining radial profiles, asymptotic behavior, and the electron localization function
\begin{equation}
    f_{\mathrm{ELF}} = \frac{1}{1 + \alpha^2},
    \label{eq:fELF}
\end{equation}
where $\alpha$ is the iso-orbital indicator in use. In Fig.~\ref{fig:iso} we show that our proposed indicators (Eq.~\ref{eq:iso-orb-OF}) yield physically correct ELF behavior in
density tail regions where the standard indicators fail. 
The full asymptotic analysis supporting these claims is given in
Appendix~\ref{app:asymptotic}.

Figure~\ref {fig:iso} shows the density profile and iso-orbital indicator on radial grid for representative atoms: Li and C. For Li, with s-orbital in valence shell, $\alpha$ and $\bar{\alpha}$ coincide throughout, both vanishing at the nucleus, rise sharply from $r\approx 1~$Bohr, indicating shell structure and then decays in the tail. This is a direct consequence of vanishing tail of density in Fig.~\ref{fig:iso}(a). In Eq.~\ref{eq:alpha_bar}, ~$\tau^{UEG}\sim \rho^{5/3}$ vanishes faster than numerator : $\tau-\tau^{vW}$, and the prefactor $\eta=10^{-3}$ in $\bar{\alpha}$ practically makes it negligible. In contrast, $\beta$ and $ \alpha_{r^{2}@OF}$ remain smooth and bounded across the full radial range since $F_{\theta}(s)\to0$ as $s\to\infty$ suppresses $\tau^{UEG}$ in the denominator, keeping the iso-orbital indicator finite even as the density vanishes. A qualitatively identical picture holds for C, where the richer shell structure, with p-orbital in outer shell, produces additional features in the interior but the same tail divergence in $\alpha$ and $\bar{\alpha}$. A much detailed analysis of asymptotic behavior is provided in Appendix.~\ref{app:asymptotic}.

The physical consequence of this divergence is most clearly seen in the electron localization function (Eq.~\ref{eq:fELF}), $f_{ELF}$, shown in Fig.~\ref{fig:ELF}. In the asymptotic tail of a finite system, electrons belong to a single exponentially decaying orbital, so the exact ELF must approach unity, the single orbital limit~\cite{Becke1990ELF}. For Li, all indicators correctly reproduce this, since $\alpha,~\bar{\alpha}\to 0$ in the tail for a diffuse s-type HOMO. For C, however the p-type HOMO produces a three-fold degenerate tail~\cite{DellaSala2015KED} in which, $\alpha$ and $~\bar{\alpha}$ diverges, driving $f_{ELF}\to0$, which is an unphysical delocalization artifact. Our modified indicators $\alpha_{r^{2}@OF}$, along with $\beta$, correctly recover $f_{ELF}\to1$ in C the tail, restoring physically consistent behavior in the low density tail region.

\begin{figure*}[htp]

    \subfigure{
        \includegraphics[width=1.0\linewidth]{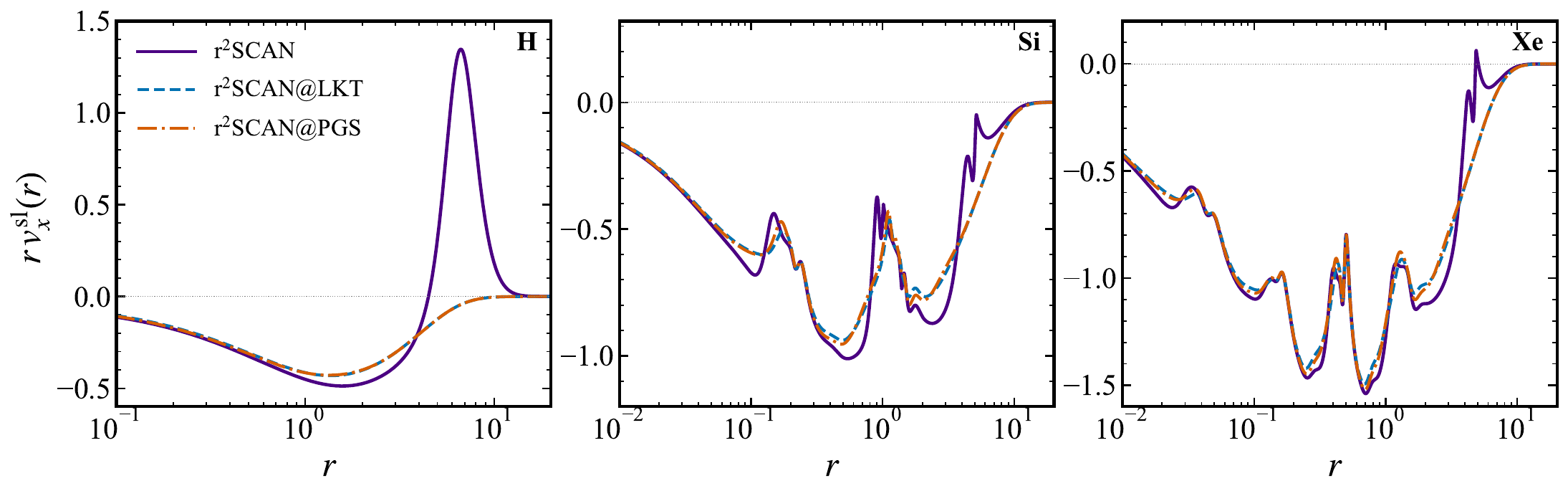}
    }\quad
    \caption{ Comparison of semilocal exchange potential $rv_x^{\mathrm{sl}}(r)$ (Ha) on a logarithmic
             radial scale (Bohr) for $r^{2}SCAN$ and our modified versions: (a) H atom (b) Si atom (b) Xe atom . The       spurious oscillations of $r^2$SCAN in the outer
             shells and tail are substantially suppressed by introducing $\alpha_{r^{2}@OF}$.}
             \label{fig:r2@OF_pot}
\end{figure*}

\begin{figure*}[htp]

    \subfigure{
        \includegraphics[width=.70\linewidth]{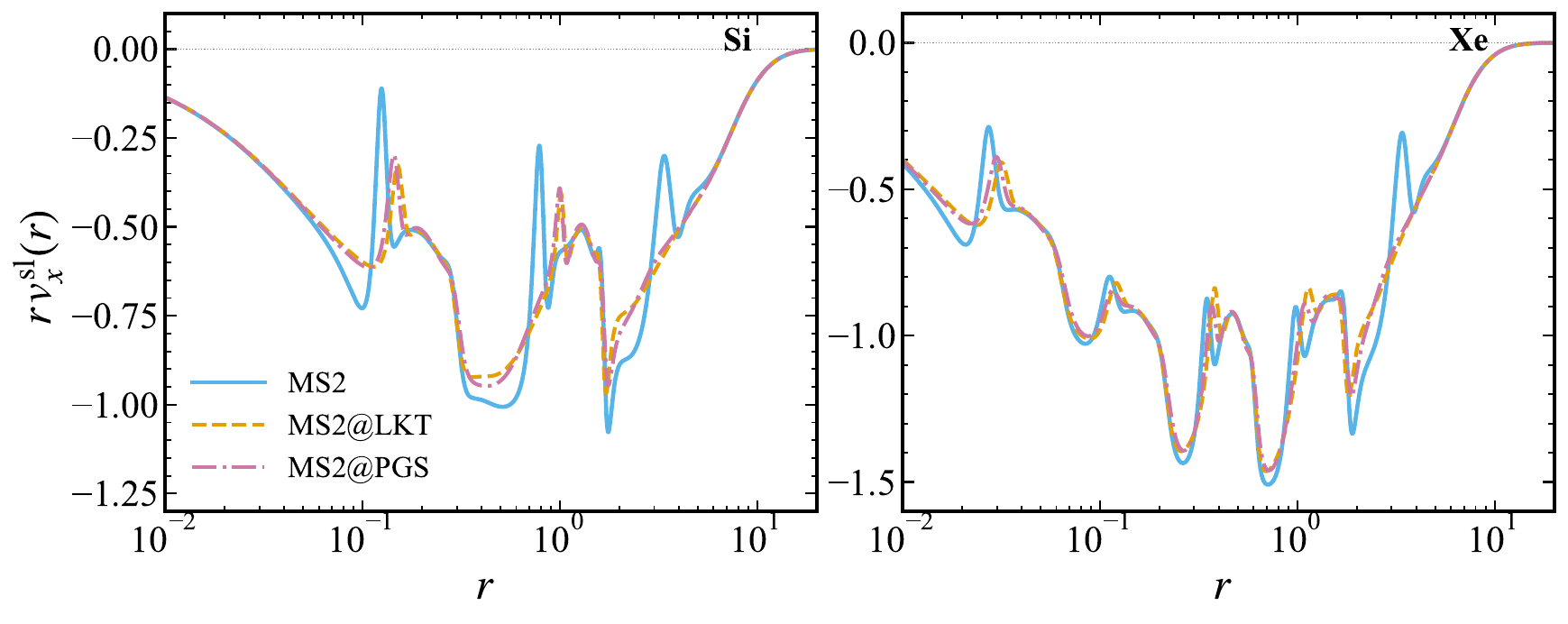}
    }\quad
    \caption{ Comparison of semilocal exchange potential $rv_x^{\mathrm{sl}}(r)$ (Ha) on a logarithmic
             radial scale (Bohr) for MS2 and our modified versions: (a) Si atom (b) Xe atom. The spurious oscillations of MS2 in the outer shells and tail are substantially suppressed by introducing $\alpha_{r^{2}@OF}$. }
             \label{fig:MS2@OF_pot}
\end{figure*}

\subsection{Exchange potential and KED  \label{sec:Results:exchange-potential}}
   
We next move on to explore how the well behaved tail of $f_{ELF}$  translates to potential of real systems. The plot of semilocal exchange potentials $rv_x^{\mathrm{sl}}(r)$ in Fig.~\ref{fig:r2@OF_pot} and Fig.~\ref{fig:MS2@OF_pot}, further ascertains our claims of correct tailing behavior. We have plotted semilocal part of exchange on radial grid using exact Hartree-Fock orbitals on three representative systems: Hydrogen atom(a),  Silicon atom(b) and Xenon atom(c) in Fig.~\ref{fig:r2@OF_pot}. Fig.~\ref{fig:MS2@OF_pot} shows plot of same quantity as in Fig.~\ref{fig:r2@OF_pot} implemented on MS2 functional. The tailing end of all the plots shows spurious oscillations in r$^2$SCAN, coming from divergent iso-orbital indicator: $\bar{\alpha}$, and adds to numerical noise. In fact, we must note here, the original SCAN functional had much wilder oscillation in potential which was reduced significantly by regularized-SCAN (rSCAN) (see Fig.4 of ~\cite{Bartok2019rSCAN})  and r$^2$SCAN (see Fig.3 of ~\cite{Furness2020r2SCAN}). In contrast, the $\alpha_{r^2@OF}$ with well behaved tail can be seen to have much smoother asymptotic potential in Fig.~\ref{fig:r2@OF_pot}. Recently, in the process of deorbitalizing r$^2$SCAN ~\cite{Francisco-Thapa_deorbita_PRB_2026}, H. Francisco et.al. in their work address the issue of unphysical bump of H-atom potential in SCAN, r$^2$SCAN, which seems to have  mitigated here with $\alpha_{r^2@OF}$. 

A natural concern is whether the smoothing of the tail region of the exchange potential introduced by $\alpha_{r^2@OF}$ removes the physically meaningful information ? 
Though in Table.~\ref{alpha_comparison} we showed our indicator distinctly identifies physically distinct regions, 
a quantitative assessment is necessary to confirm  that improvement is not achieved at the expense of essential features.
However, smoothness alone is not a sufficient criteria for functional quality; for example , LDA yields exceptionally smooth potentials while failing to capture important physical characteristics for real systems. Therefore, any modification that improves numerical behavior must also be shown to preserve the underlying physical content.  
 Is $F_\theta$ modification  removes physical content from the iso-orbital indicator in bonding regions. The answer is no, because in regions of high electron density,  
where $s$ is moderate, $F_\theta(s) \approx 1$ and the modified denominator $\tau^{\mathrm{UEG}}F_\theta + \tau^{\mathrm{vW}}$ 
regains  approximately $\tau^{\mathrm{UEG}} + \tau^{\mathrm{vW}}$, 
essentially unchanged from the original.

We know that for Si, the valence configuration 3s$^2$,3p$^2$, and its 3s--3p valence and tail region,where the electron density decays and $s$ becomes large. Figure~\ref{fig:r2@OF_pot} shows that in the core region ($r \lesssim 0.2$~Bohr) all three functionals agree perfectly, confirming again that the $F_{\theta}$ modification is inert where it should be. In the valence region ($r \approx 0.5$--$2$~Bohr), r$^2$SCAN produces a pronounced dip and oscillatory features that are visibly suppressed in  $\alpha_{r^{2}@OF}$ variants. Next Fig.~\ref{fig:MS2@OF_pot} supports the observation of Fig.~\ref{fig:r2@OF_pot} in every way confirming the generic improvement in iso-orbital description can be plugged in any meta-GGA functional. The use of $\alpha_{r^2@OF}$ in both r$^2$SCAN and MS2 with PGS and LKT substantially suppress these oscillations, producing a monotonically decaying, smooth potential throughout the tail region. Though MS2 uses a different interpolation function and exchange enhancement factor from r$^2$SCAN and yet  the oscillation are suppressed which is a consequence of the corrected iso-orbital indicator rather than any feature specific to one functional form. The practical consequences of such residual oscillations are well documented, they cause instabilities in pseudopotential generation~\cite{Bartok2019rSCAN}, produce pathological grid-convergence behavior in self-consistent calculations~\cite{Furness2020r2SCAN, review_KED_Della_Sala_2016}, and amplify into much larger errors in second functional derivatives i.e. kernels of linear response, required for TDDFT kernels~\cite{Furness2020r2SCAN}. The further smoothing achieved by $\alpha_{r^{2}@OF}$ therefore carries direct numerical benefits beyond the present accuracy assessment.

\begin{figure*}
    \centering
    \includegraphics[scale=0.40]{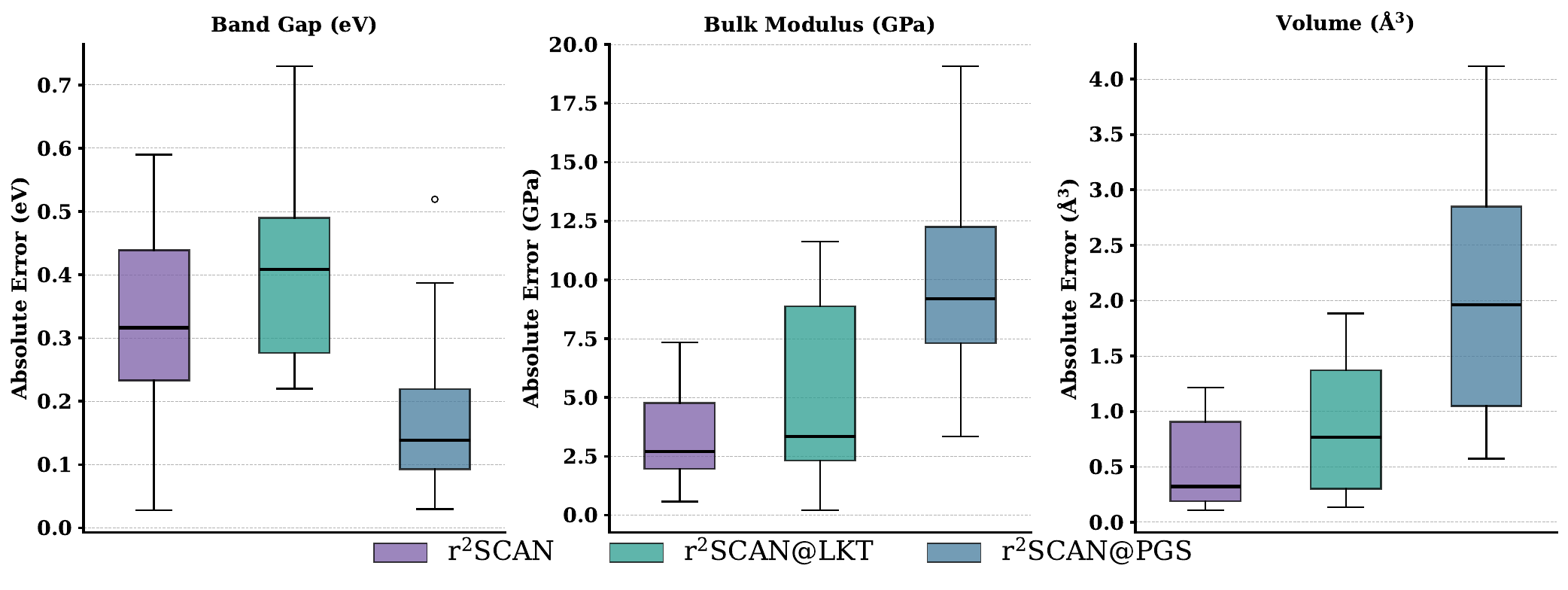}
    \caption{Boxplots of absolute errors in (left) band gap (eV), (center) bulk modulus(GPa), and (right) equilibrium volume (\AA$^3$) for the cubic semiconductor test set (Si, AlAs, AlP, AlSb, GaAs, GaP, GaSb, InSb, InP, InAs) computed with r$^2$SCAN, r$^2$SCAN@LKT,r$^2$SCAN@PGS.}
    \label{fig:SC_R2S}
\end{figure*}

\begin{figure*}
    \centering
    \includegraphics[scale=0.40]{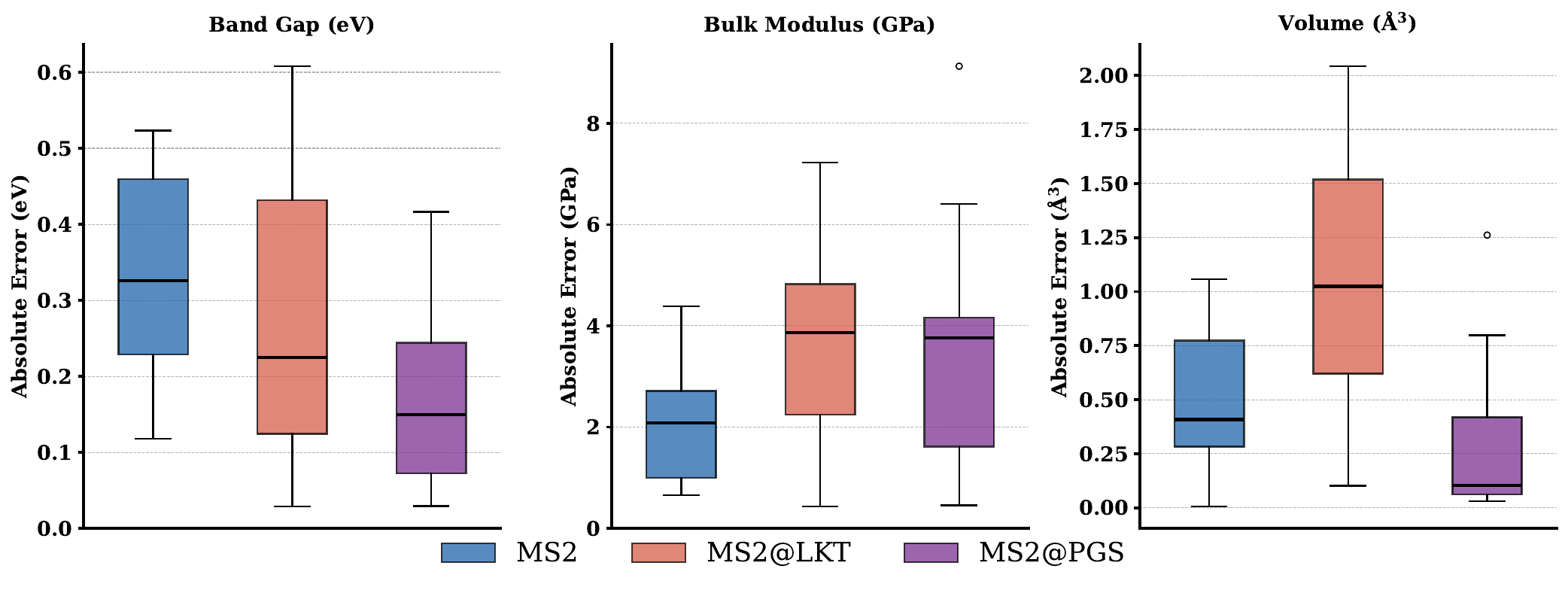}
    \caption{Same as in Fig.~\ref{fig:SC_R2S} for MS2, MS2@LKT, and MS2@PGS on the same semiconductor test set.}
    \label{fig:boxplot_MS2}
\end{figure*}

\subsection{Application on solids \label{sec:Results:Solids} }

\begin{table}[htbp]
\centering
\caption{Mean absolute errors (MAE) and root mean square errors (RMSE) for band gap (B.G)
         (eV), equilibrium volume (V$_0$) in \AA$^3$, and bulk modulus (B$_0$) in GPa on the semiconductor test set, 
         which were computed against the experimental reference values are taken from Ref~\cite{Volume} ($V_0$), Ref~\cite{Bandgap2}($B_0$), and
         Ref~\cite{Bandgap}(B.G).}
\label{tab:mae_rmse}
\setlength{\tabcolsep}{10pt}
\begin{tabular}{lcccc}
\hline\hline
Functional & B.G (eV) & V$_0$ (\AA$^3$) & B$_0$ (GPa) \\
\hline
\multicolumn{4}{l}{\textit{MAE}} \\
r$^2$SCAN          & 0.321 & 0.509 &  3.335 \\
r$^2$SCAN@LKT      & 0.409 & 0.861 &  5.077 \\
r$^2$SCAN@PGS      & 0.189 & 1.992 & 10.462 \\
MS2             & 0.334 & 0.493 & 2.166 \\
MS2@LKT         & 0.264 & 1.039 & 3.645 \\
MS2@PGS         & 0.171 & 0.327 & 3.621 \\
\hline
\multicolumn{4}{l}{\textit{RMSE}} \\
r$^2$SCAN          & 0.359 & 0.659 &  3.969 \\
r$^2$SCAN@LKT      & 0.436 & 1.042 & 6.409 \\
r$^2$SCAN@PGS      & 0.238 & 2.282 & 11.510 \\
MS2             & 0.364 & 0.599 & 2.511 \\
MS2@LKT         & 0.328 & 1.207 & 4.167 \\
MS2@PGS         & 0.207 & 0.509 & 4.403 \\
\hline\hline
\end{tabular}

\end{table}

\begin{figure}
    \centering
    \includegraphics[width=1.0\linewidth]{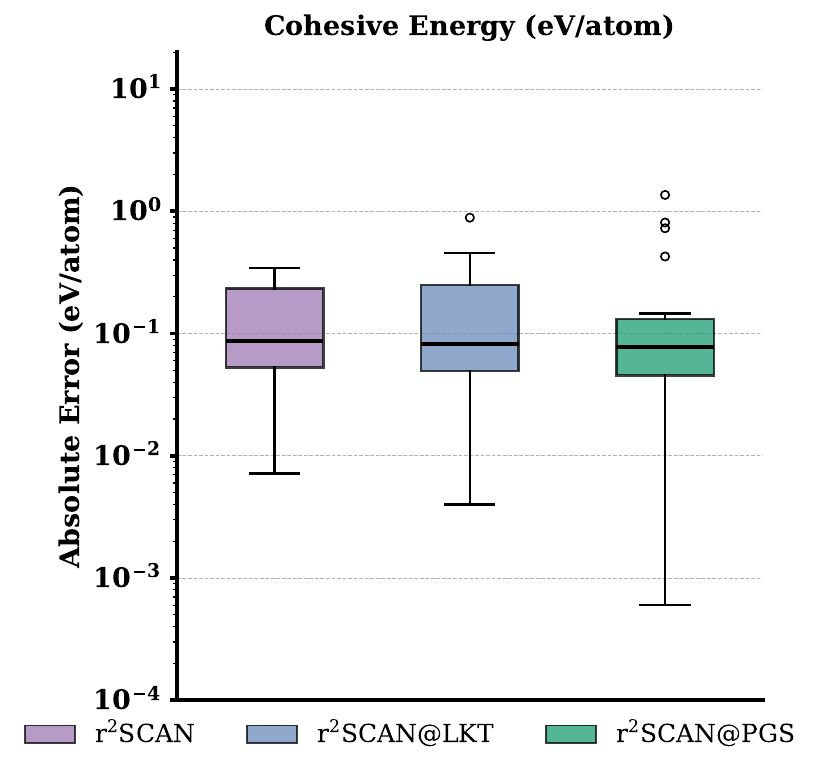}
    \caption{Box plots of absolute errors in cohesive energy (eV/atom) for r$^2$SCAN,
             r$^2$SCAN@LKT, and r$^2$SCAN@PGS on the 23 material benchmark (GaAs, Al, and
             MgO excluded as modified functionals did not reach the required SCF convergence threshold).  r$^2$SCAN@PGS achieves the lowest median error and
             the tightest interquartile range of the three functionals, demonstrating
             that the tail region regularization does not degrade.  Outliers at 0.89 eV/atom for r$^2$SCAN@LKT for C and above $1$~eV/atom for r$^2$SCAN@PGS correspond to noble and late transition metals (Rh, Pd, Ag).}
    \label{fig:cohesive}
\end{figure}



Fig.~\ref{fig:SC_R2S},~\ref{fig:boxplot_MS2} and   Table~\ref{tab:mae_rmse} presents the calculation of band gap,volume,bulk modulus for the ten member of semiconductor test set, for more details see supplementary~\cite{support}. Band gap accuracy requires a dependence of KED in the exchange-correlation potential ~\cite{Tran2009TBmBJ,Tran_BG_benchmark_solids,LAK_Lebeda_Kummel_PRL_2024}. 
Analyzing the mean absolute error (MAE) of band gap in Table~\ref{tab:mae_rmse}, we can conclude 
$\alpha_{r2@OF}$ with PGS has improved it's both parent functionals r$^2$SCAN by 41.1\% and a 48.8\% in MS2. Whereas with LKT enhancement factor deteriorates for r$^2$SCAN@LKT by 27.4\% and improves MS2 by 21\%. This observation can be physically understood. In the interstitial region of a typical III--V semiconductor, the reduced gradient reaches $s \approx 1$--$3$~\cite{Tran_BG_benchmark_solids}. Now, the $F_{\theta}(s)$ screens the TF part in PGS in this very range in Fig.~\ref{fig:F_theta_enhance}, which is more suitable approximation for semiconductors. Besides, r$^2$SCAN is a much complicated functional than MS2 with tightly fitted parameters adjusted for $\bar{\alpha}$, which we did not retune for $\alpha_{r^2@OF}$ in this work.

For bulk modulus, $r^2$SCAN achieves lowest MAE of $3.335\,\mathrm{GPa}$ and equilibrium volume of MAE of $0.509\,\text{\AA}^3$. Both $r^2$SCAN@LKT and $r^2$SCAN@PGS degrade these properties by margin $\sim 5.077\,\mathrm{GPa}, \sim 10.462\,\mathrm{GPa}$ for $B_0$ and $\sim0.86\,\text{\AA}^3, \sim1.992\,\text{\AA}^3$ for $V_0$ respectively. This degradation structural properties stems from deviation from linear response of tightly optimized $r^2$SCAN. The response can be restored by tuning the parameter $\mu$ in PGS (or $a$ in LKT). However, we here stick to provide alternative form of iso-orbital rather than improving performance of any particular functional. This claim is supported by the result of MS2 variants of functionals in Fig.~\ref{fig:boxplot_MS2}, this degradation is much less in $B_0$, in fact the $V_0$ improves for MS2@PGS by $33.7\%$, the best volume accuracy among all six functionals studied. Moreover, the bulk modulus is a second derivative of the energy with respect to volume, so any residual oscillations in $v_{\mathrm{XC}}$ are magnified in this derivative. We consciously refrain from reparametrizing $\mu$ and $a$ to recover structural accuracy in order to preserve the transferability of the approach.

\begin{table}[htbp]
\centering
\caption{Cohesive energies (eV/atom) computed with r$^2$SCAN, r$^2$SCAN@LKT, and r$^2$SCAN@PGS
         compared against experimental reference ~\cite{OFR2} values for a 23-material benchmark set.
         GaAs, Al, and MgO are excluded as the isolated open shell atomic reference
         energies for the modified functionals did not reach the required SCF convergence
         threshold.}
\label{tab:cohesive}
\setlength{\tabcolsep}{4pt}
\begin{tabular}{lcccc}
\hline\hline
Material & Ref.~\cite{OFR2} & r$^2$SCAN & r$^2$SCAN@LKT & r$^2$SCAN@PGS \\
\hline
\multicolumn{5}{l}{\textit{Alkali metals}} \\
Li   & 1.670 & 1.677 & 1.732 & 1.708 \\
Na   & 1.120 & 1.075 & 1.130 & 1.121 \\
K    & 0.940 & 0.846 & 0.904 & 0.910 \\
Rb   & 0.860 & 0.747 & 0.805 & 0.810 \\
Cs   & 0.810 & 0.681 & 0.763 & 0.754 \\
\hline
\multicolumn{5}{l}{\textit{Alkaline earth metals}} \\
Ca   & 1.870 & 2.059 & 1.974 & 1.919 \\
Sr   & 1.730 & 1.785 & 1.623 & 1.584 \\
Ba   & 1.910 & 1.991 & 2.000 & 1.830 \\
\hline
\multicolumn{5}{l}{\textit{Transition metals}} \\
Cu   & 3.510 & 3.854 & 3.436 & 3.083 \\
Rh   & 5.780 & 5.493 & 5.732 & 4.419 \\
Pd   & 3.930 & 4.162 & 3.880 & 3.203 \\
Ag   & 2.960 & 2.882 & 2.717 & 2.151 \\
\hline
\multicolumn{5}{l}{\textit{Covalent solids}} \\
C    & 7.550 & 7.890 & 8.437 & 7.474 \\
SiC  & 6.480 & 6.766 & 6.936 & 6.353 \\
Si   & 4.680 & 4.919 & 4.882 & 4.581 \\
Ge   & 3.890 & 3.878 & 4.199 & 3.914 \\
\hline
\multicolumn{5}{l}{\textit{Ionic solids}} \\
LiF  & 4.460 & 4.451 & 4.741 & 4.490 \\
LiCl & 3.590 & 3.529 & 3.648 & 3.505 \\
NaF  & 3.970 & 4.010 & 4.238 & 4.018 \\
NaCl & 3.340 & 3.279 & 3.344 & 3.255 \\
\hline
MAE  &       & 0.135 & 0.170 & 0.217 \\
RMSE &       & 0.175 & 0.264 & 0.407 \\
ME   &       & $+$0.046 & $+$0.104 & $-$0.198 \\
\hline\hline
\end{tabular}
\end{table}

The cohesive energy is dominated by the XC energy in the
high density bonding region~\cite{Sun2015SCAN}, providing a
critical test of whether the $F_\theta$ correction inadvertently
disrupts bonding region energetics while correcting the tail.

Table~\ref{tab:cohesive} and Fig.~\ref{fig:cohesive} 
show results
for the 23 material benchmark, supporting data can be found in suplementary~\cite{support} data. Despite having the largest MAE
(0.217~eV/atom) and RMSE (0.407~eV/atom), r$^2$SCAN@PGS achieves
the lowest median absolute error and tightest interquartile range
of the three functionals. The contradiction is resolved by the
error distribution: the bulk of the benchmark is described more
accurately by r$^2$SCAN@PGS than by r$^2$SCAN, but three outliers
, the noble and late transition metals Rh, Pd, and Ag, where PGS
screening reaches into the $d$-electron bonding region, pull the
MAE and RMSE upward disproportionately.

The mean error reveals the most physically consequential trend,
r$^2$SCAN slightly overestimates cohesive energies
(ME $= +0.046$~eV/atom)~\cite{Furness2020r2SCAN};
r$^2$SCAN@LKT increases this to ME $= +0.104$~eV/atom, while
r$^2$SCAN@PGS reverses the sign to ME $= -0.198$~eV/atom. This
sign reversal reflects PGS's stronger reduction of the XC energy
at intermediate $s$, yielding a systematically less bound solid.
The crossover from overestimation to underestimation between LKT
and PGS directly constrains the acceptable range of $F_\theta$
decay rates in future functional development, the optimal
parametrization lies between the two forms.

Material class trends in Table~\ref{tab:cohesive} are consistent
with this picture, alkali metals (large radius, diffuse $s$
electrons, large interstitial $s$) benefit most from $F_\theta$
screening; transition metals are most sensitive to PGS's aggressive
intermediate $s$ decay, ionic and covalent solids are largely
unaffected.

\section{Conclusion}
\label{sec:conclusion}

From the above results, we can construct a physically motivated narrative that connects the
mathematical form of $F_{\theta}$ for screened iso-orbital indicator to improvements across a hierarchy of properties. The key findings are summarized below.

The divergence of $\alpha$ in the atomic tail is a fundamental issue with measurable
consequences at every level of analysis.  It produces an unphysical ELF in the carbon
tail , oscillations in the Xe exchange potential
, and systematic underestimation of semiconductor band gaps
.  The $F_{\theta}$ screening eliminates this divergence
by suppressing $\tau_{\mathrm{UEG}}$ proportionally to the reduced gradient $s$,
providing a correction that is both targeted acting only where $s$ is large, leaving
the bonding region undisturbed and transferable effective across r$^2$SCAN and MS2 parent
functionals with different mathematical constructions.

The PGS enhancement factor, with its faster Gaussian decay, provides a more effective
correction than LKT across both parent functionals, reducing semiconductor band gap MAE
by 41.1\% (r$^2$SCAN) and 48.8\% (MS2) and 33.7\% improvement in volume in MS2@PGS while increasing cohesive energy MAE by at most
0.08~eV/atom.  The LKT factor, with its slower decay rate, is effective
only in the extreme $s$ tail and fails to correct the intermediate $s$ interstitial
regime responsible for band gap errors.  The slight increase in bulk modulus errors for
r$^2$SCAN is physically expected given that functional's tight optimization for structural
properties, and is not a fundamental limitation of the approach.

These results provide a clear physical foundation and quantitative motivation for the
construction of a new meta-GGA functional in which the $F_{\theta}$ decay rate is
optimized simultaneously with the exchange-correlation enhancement factor parameters.
Such a functional would balance the tail region correction against the bonding region
constraint, recovering r$^2$SCAN's structural accuracy while keeping  the band gap
improvement.

\section*{Acknowledgments}
The authors will like to thank the computational facilities provided by National Institute of Science Education and Research(NISER), India.

\section*{Data Availability}
We confirm that the data supporting the findings of this study are available within the article and its supplementary materials ~\cite{support}.

\appendix
\section{Asymptotic regions analysis of iso-orbital indicators}
\label{app:asymptotic}

We here show the behavior of various iso-orbitals indicators in the important asymptotic limit. The deductions are based on the exact shell decomposition of Kohn-Sham KED ~\cite{DellaSala2015KED}:
\begin{equation}
    \tau_{nl}^{\mathrm{KS}} = \tau^{vW}[\rho_{nl}] + \frac{l(l+1)}{2}\frac{\rho_{nl}}{r^2}
    \label{eq:app:shell-decomp}
\end{equation}
The four indicators we consider in main text as defined in Eq.(~\ref{eq:alpha}),(\ref{eq:alpha_bar}),(\ref{eq:beta}) and (\ref{eq:iso-orb-OF}), are defined as:
\begin{align}
    \alpha &= \frac{\tau - \tau^{\mathrm{vW}}}{\tau^{\mathrm{UEG}}},
     \nonumber \\
    \bar{\alpha} &= \frac{\tau - \tau^{\mathrm{vW}}}{\tau^{\mathrm{UEG}} + \eta\,\tau^{\mathrm{vW}}},
    \quad \eta = 10^{-3},
    \nonumber  \\
    \beta &= \frac{\tau - \tau^{\mathrm{vW}}}{ \tau + \tau^{\mathrm{UEG}} },
    \nonumber  \\
    \alpha_{r^2@\mathrm{OF}} &= \frac{\tau - \tau^{\mathrm{vW}}}{\tau^{\mathrm{UEG}} F_\theta(s) + \tau^{\mathrm{vW}}}, \nonumber 
\label{eq:app:alphaOF}
\end{align}


\subsection{Single orbital limit: Tail region of $s$-type HOMO ($l = 0$)}

In the asymptotic tail, only the HOMO contributes, thus can be identified as single orbital region. The KS KED equals the von Weizs\"{a}cker KED exactly: $\tau = \frac{1}{2}|\nabla\phi|^2 = \frac{|\nabla\rho|^2}{8\rho} = \tau^{\mathrm{vW}}$.

For an $s$-type HOMO, $l = 0$, and from
Eq.~\eqref{eq:app:shell-decomp}:
\begin{equation}
\tau_{n0}^{\mathrm{KS}} = \tau^{\mathrm{vW}}[\rho_{n0}]
+ \frac{0 \cdot 1}{2}\frac{\rho_{n0}}{r^2}
= \tau^{\mathrm{vW}}[\rho_{n0}].
\end{equation}
Therefore $\tau = \tau^{\mathrm{vW}}$  in each $s$-shell,
and the numerator of every indicator vanishes:
\begin{equation}
\tau - \tau^{\mathrm{vW}} = 0  .
\end{equation}
Since the numerator is zero:
\begin{equation}
\alpha = \bar{\alpha} = \beta = \alpha_{r^2\mathrm{OF}} = 0.
\end{equation}
All indicators correctly identify the single orbital character of the $s$ type tail~\cite{DellaSala2015KED,MS2_beta_Furness2019_isoorbital}.


\subsection{Slowly varying density limit}
In the slowly varying limit, the reduced gradient $s \to 0$, which implies:
\begin{equation}
\tau^{\mathrm{vW}} = \frac{|\nabla\rho|^2}{8\rho} \to 0,
\qquad
\tau \approx \tau^{\mathrm{UEG}}.
\end{equation}
The second condition is the gradient expansion result for the KS KED~\cite{Sun2015SCAN}. Since $s \to 0$, both enhancement factors recover $F_\theta(0) = 1$. Therefore,
\begin{equation}
\alpha \approx \bar{\alpha} \approx \alpha_{r^2\mathrm{OF}} \approx 1,
\qquad \beta \approx \tfrac{1}{2}.
\end{equation}
The slowly-varying limit $\approx 1$ is preserved by $\alpha_{r^2\mathrm{OF}}$
because $F_\theta(0) = 1$ is a design constraint of both LKT and PGS,
not an empirical choice. This is the key advantage over $\beta$.


\subsection{Tail region: $p$-type HOMO ($l = 1$)}
This is the critical case. For a $p$-type
HOMO, $l = 1$, and from Eq.~\eqref{eq:app:shell-decomp}:
\begin{equation}
\tau_{n1}^{\mathrm{KS}} = \tau^W[\rho_{n1}]
+ \frac{1 \cdot 2}{2}\frac{\rho_{n1}}{r^2}
= \tau^W[\rho_{n1}] + \frac{\rho_{n1}}{r^2}.
\label{eq:app:p-type-KED}
\end{equation}
In the asymptotic region only the HOMO contributes, so the Pauli
KED is:
\begin{equation}
\tau - \tau^{\mathrm{vW}} = \frac{\rho_{n1}}{r^2} > 0,
\label{eq:app:pauli-ptail}
\end{equation}
which is strictly positive even as $\rho \to 0$. This is the fundamental
reason why the p-type tail is qualitatively different from the s-type
tail~\cite{DellaSala2015KED}.
 The exponential decay of the HOMO $\rho \sim \,e^{-2\kappa r}, \qquad
\kappa = \sqrt{-2\epsilon_{\mathrm{HOMO}}} > 0$ .
From this, all relevant quantities scale as:
\begin{align}
\tau^{\mathrm{vW}} &= \frac{|\nabla\rho|^2}{8\rho}
\approx \frac{(2\kappa\rho)^2}{8\rho}
= \frac{\kappa^2}{2}\,\rho,
\label{eq:app:tauvW-ptail}\\
\tau - \tau^{\mathrm{vW}} &= \frac{\rho}{r^2}
\quad (\text{from Eq.~\eqref{eq:app:pauli-ptail}}),
\label{eq:app:numerator-ptail}\\
\tau^{\mathrm{UEG}} & \propto \rho^{5/3} .
\label{eq:app:tauUEG-ptail}
\end{align}

Therefore the approximation: $\tau^{\mathrm{UEG}} F_\theta + \tau^{\mathrm{vW}}\approx \frac{\kappa^2}{2}\,\rho$
is exact in the $\rho \to 0$ limit.
We now go through each indicator separately to examine the behavior in the tail region.
\begin{equation}
\alpha = \frac{\tau - \tau^{\mathrm{vW}}}{\tau^{\mathrm{UEG}}}
\approx \frac{\rho/r^2}{\rho^{5/3}}
= \frac{1}{r^2}\,\rho^{-2/3} \to \infty,
\label{eq:app:alpha-diverge}
\end{equation}
since $\rho^{-2/3} \to \infty$ faster than $r^{-2} \to 0$.  

\begin{equation}
\bar{\alpha}
= \frac{\tau - \tau^{\mathrm{vW}}}
{\tau^{\mathrm{UEG}} + \eta\,\tau^{\mathrm{vW}}}
\approx \frac{\rho/r^2}{\rho^{5/3} + \eta\,(\kappa^2/2)\rho}.
\label{eq:app:alphabar-ptail-1}
\end{equation}
As $\rho \to 0$, $\rho^{5/3} \ll \rho$, so the denominator
$\approx \eta\,(\kappa^2/2)\,\rho$:
\begin{equation}
\bar{\alpha} \approx \frac{\rho/r^2}{\eta\,(\kappa^2/2)\,\rho}
= \frac{2}{\eta\kappa^2 r^2}.
\label{eq:app:alphabar-ptail-2}
\end{equation}
Although $\bar{\alpha} \to 0$ formally as $r \to \infty$, at
intermediate $r$ values it passes through:
\begin{equation}
\bar{\alpha} \sim \frac{2}{\eta\kappa^2 r^2}
= \mathcal{O}(1/\eta) \sim 10^2\text{-}10^3
\quad \text{for } \eta = 10^{-3},
\label{eq:app:alphabar-large}
\end{equation}
which is large enough to drive the exchange enhancement factor into a
physically incorrect regime and produce the spurious oscillations seen
in Figs.~\ref{fig:iso}~\ref{fig:ELF}and~\ref{fig:r2@OF_pot} see Ref.~\cite{Furness2020r2SCAN}.

\begin{equation}
\beta = \frac{\tau - \tau^{\mathrm{vW}}}{\tau + \tau^{\mathrm{UEG}}}.
\end{equation}
In the p-type tail, $\tau \approx \tau^{\mathrm{vW}} + \rho/r^2$
(from Eq.~\eqref{eq:app:p-type-KED}), and both numerator and denominator
scale as $\rho$:
\begin{equation}
\beta \approx \frac{\rho/r^2}
{(\kappa^2/2)\rho + \rho/r^2 + \rho^{5/3}}
\approx  \frac{2}{\kappa^2 r^2} \to 0.
\label{eq:app:beta-ptail}
\end{equation}
$\beta$ goes to zero smoothly, giving $f_{\mathrm{ELF}} \to 1$, which is
physically correct~\cite{MS2_beta_Furness2019_isoorbital}. However, at intermediate
distances $\beta$ approaches a finite saturation value $< 1$ rather than
$\approx 1$ in the bonding region, compressing the bonding-environment
identification~\cite{MS2_beta_Furness2019_isoorbital}.

\begin{equation}
\alpha_{r^2\mathrm{OF}}
\approx \frac{\rho/r^2}{(\kappa^2/2)\,\rho}
= \frac{2}{\kappa^2 r^2} \to 0.
\label{eq:app:alphaOF-ptail}
\end{equation}
The indicator vanishes as $r^{-2}$ in the p-type tail. Therefore as $r \to \infty$:
\begin{equation}
f_{\mathrm{ELF}} = \frac{1}{1 + \alpha_{r^2\mathrm{OF}}^2}
= \frac{1}{1 + 4/(\kappa^4 r^4)} \to 1
\label{eq:app:ELF-ptail}
\end{equation}
correctly recovering the single-orbital limit of the
ELF~\cite{Becke1990ELF}. This is in sharp contrast to $\alpha$ and $\bar{\alpha}$, for
which $f_{\mathrm{ELF}} \to 0$ in the p-type tail, producing the
unphysical delocalization artifact seen for the C atom in
Fig.~\ref{fig:ELF}.

\subsection{Noncovalent density overlap region}

In the low-density interstitial region between two weakly
interacting closed-shell systems, the important conditions
are~\cite{MS2_beta_Furness2019_isoorbital}:
\begin{equation}
\rho \to 0, \qquad \tau^{\mathrm{vW}} = 0
\;\text{at bond centers},
\end{equation}
with $\tau^{\mathrm{UEG}} \propto \rho^{5/3}$ and
$\tau - \tau^{\mathrm{vW}} = \tau \propto \rho$~ (~\cite{MS2_beta_Furness2019_isoorbital}.
Since $\rho^{5/3}$ vanishes faster than $\rho$:
\begin{align}
\alpha &\sim \frac{\rho}{\rho^{5/3}}
= \rho^{-2/3} \to \infty,
\label{eq:app:noncov-alpha}\\
\bar{\alpha} &\sim \rho^{-2/3} \to \infty,
\label{eq:app:noncov-alphabar}
\end{align}
where for $\bar{\alpha}$ the $\eta$ floor vanishes since
$\tau^{\mathrm{vW}} = 0$ at the bond center, leaving
the denominator $\approx \tau^{\mathrm{UEG}} \propto
\rho^{5/3}$~\cite{Furness2020r2SCAN}. For $\beta$, both
numerator and denominator scale as $\rho$, giving a finite
value~\cite{MS2_beta_Furness2019_isoorbital}:
\begin{equation}
\beta \sim \frac{\rho}{\rho + \rho^{5/3}}
\approx \text{finite},\quad \tfrac{1}{2} \ll \beta < 1.
\label{eq:app:noncov-beta}
\end{equation}
For $\alpha_{r^2\mathrm{OF}}$, with $\tau^{\mathrm{vW}} = 0$
and $s \propto \nabla \rho \to 0$, the denominator
reduces to $\tau^{\mathrm{UEG}} F_\theta$. As $s \to 0$, the two enhancement factors decay as:
\begin{align}
F_\theta^{\mathrm{PGS}} &= e^{-\mu s^2}
\approx 1,
\label{eq:app:noncov-FPGS}\\
F_\theta^{\mathrm{LKT}} &= \frac{1}{\cosh(as)}
\
\approx 1,
\label{eq:app:noncov-FLKT}
\end{align}
The resulting indicators are:
\begin{align}
\alpha_{r^2\mathrm{OF}}^{\mathrm{PGS}}
&\sim \frac{\rho}{\rho^{5/3}}
= \rho^{-2/3} \to \infty,
\label{eq:app:noncov-alphaOF-PGS}\\
\alpha_{r^2\mathrm{OF}}^{\mathrm{LKT}}
&\sim \frac{\rho}{\rho^{5/3}}
= \rho^{-2/3} \to \infty.
\label{eq:app:noncov-alphaOF-LKT}
\end{align}
Both formally diverge at the strict bond center. And it behaves similar to $\alpha$ and $\bar{\alpha}$.

\twocolumngrid
\bibliography{reference}
\bibliographystyle{apsrev}

\end{document}


\title{Physically motivated iso-orbital indicator for meta-GGA exchange functionals}
\author{Jeet Sharma}
\email{jeet.sharma@niser.ac.in}
\affiliation{School of Physical Sciences, National Institute of Science Education and Research, An OCC of Homi Bhabha National Institute, Jatni 752050, India}
\author{Abhishek Bhattacharjee}
\affiliation{School of Physical Sciences, National Institute of Science Education and Research, An OCC of Homi Bhabha National Institute, Jatni 752050, India}
\author{Bikash Patra}
\affiliation{Department of Physical and Applied Sciences
, Indian Institute of Information Technology Surat, Kholvad campus, Kamrej 394190, Surat (Gujarat), India}

 \author{Prasanjit Samal}
 \affiliation{School of Physical Sciences, National Institute of Science Education and Research, An OCC of Homi Bhabha National Institute, Jatni 752050, India}

 \maketitle

\begin{*table}
{\fontsize{10}{12}\selectfont
\setlength{\LTcapwidth}{\textwidth}
\setlength{\LTleft}{0pt}
\setlength{\LTright}{0pt}
\begin{longtable}{@{\extracolsep{\fill}}ccccc@{}}
\caption{\label{tab:semi}
Equilibrium volume per atom ($V_0$, in \AA$^3$), bulk modulus ($B_0$, in GPa),
and band gap (B.G., in eV) of semiconductors computed with r$^2$SCAN and MS2,
along with their variants employing the modified iso-orbital indicators
(@PGS, @LKT). Experimental reference values are taken from
Ref.~\cite{Volume} ($V_0$), Ref.~\cite{Bandgap2} ($B_0$), and
Refs.~\cite{Bandgap,Bandgap2} (B.G.).}\\
\hline\hline
Elements & Functional & $V_0$ (\AA$^3$) & $B_0$ (GPa) & B.G. (eV) \\
\hline
\endfirsthead

\caption[]{(Continued.)}\\
\hline\hline
Elements & Functional & $V_0$ (\AA$^3$ /atom) & $B_0$ (GPa) & B.G. (eV) \\
\hline
\endhead

\hline
\multicolumn{5}{r}{\textit{Continued on next page}}\\
\endfoot

\hline\hline
\endlastfoot

CD-Si & Ref.                & 20.012 & 100.80  & 1.17  \\
      & r$^2$SCAN           & 20.129 & 95.536  & 0.77  \\
      & r$^2$SCAN@PGS       & 20.611 & 88.883  & 1.32  \\
      & r$^2$SCAN@LKT       & 20.323 & 90.519  & 0.93  \\
      & MS2                 & 20.006 & 100.150 & 1.05  \\
      & MS2@PGS             & 19.951 & 104.321 & 1.58  \\
      & MS2@LKT             & 20.112 & 97.233  & 0.95  \\
\hline
AlAs  & Ref.                & 22.641 & 75.00   & 2.23  \\
      & r$^2$SCAN           & 22.868 & 76.934  & 1.78  \\
      & r$^2$SCAN@PGS       & 23.642 & 67.910  & 2.11  \\
      & r$^2$SCAN@LKT       & 22.942 & 72.271  & 1.82  \\
      & MS2                 & 22.905 & 76.270  & 1.77  \\
      & MS2@PGS             & 22.671 & 79.203  & 2.33  \\
      & MS2@LKT             & 23.196 & 74.196  & 1.73  \\
\hline
AlP   & Ref.                & 20.346 & 87.40   & 2.50  \\
      & r$^2$SCAN           & 20.460 & 90.473  & 1.91  \\
      & r$^2$SCAN@PGS       & 20.920 & 84.064  & 2.38  \\
      & r$^2$SCAN@LKT       & 20.482 & 87.611  & 1.99  \\
      & MS2                 & 20.428 & 91.546  & 1.97  \\
      & MS2@PGS             & 20.377 & 91.425  & 2.61  \\
      & MS2@LKT             & 20.592 & 89.200  & 1.89  \\
\hline
AlSb  & Ref.                & 28.877 & 55.10   & 1.69  \\
      & r$^2$SCAN           & 29.327 & 61.482  & 1.45  \\
      & r$^2$SCAN@PGS       & 30.921 & 48.897  & 1.66  \\
      & r$^2$SCAN@LKT       & 29.733 & 53.970  & 1.47  \\
      & MS2                 & 29.452 & 56.630  & 1.47  \\
      & MS2@PGS             & 28.963 & 60.865  & 1.87  \\
      & MS2@LKT             & 29.922 & 54.914  & 1.44  \\
\hline
GaAs  & Ref.                & 22.521 & 76.70   & 1.52  \\
      & r$^2$SCAN           & 22.805 & 74.618  & 1.08  \\
      & r$^2$SCAN@PGS       & 24.402 & 57.631  & 1.00  \\
      & r$^2$SCAN@LKT       & 23.201 & 65.087  & 0.79  \\
      & MS2                 & 22.938 & 72.316  & 1.00  \\
      & MS2@PGS             & 22.640 & 74.934  & 1.54  \\
      & MS2@LKT             & 23.524 & 69.476  & 1.36  \\
\hline
GaP   & Ref.                & 20.212 & 89.60   & 2.35  \\
      & r$^2$SCAN           & 20.397 & 88.518  & 1.88  \\
      & r$^2$SCAN@PGS       & 21.412 & 71.342  & 2.27  \\
      & r$^2$SCAN@LKT       & 20.518 & 79.821  & 1.95  \\
      & MS2                 & 20.610 & 88.849  & 1.88  \\
      & MS2@PGS             & 20.638 & 85.588  & 2.40  \\
      & MS2@LKT             & 21.024 & 85.534  & 1.83  \\
\hline
GaSb  & Ref.                & 28.316 & 51.51   & 0.82  \\
      & r$^2$SCAN           & 28.685 & 58.862  & 0.57  \\
      & r$^2$SCAN@PGS       & 31.270 & 43.565  & 0.43  \\
      & r$^2$SCAN@LKT       & 29.495 & 49.348  & 0.28  \\
      & MS2                 & 28.650 & 53.514  & 0.45  \\
      & MS2@PGS             & 27.921 & 60.865  & 0.88  \\
      & MS2@LKT             & 29.450 & 51.089  & 0.78  \\
\hline
InAs  & Ref.                & 27.735 & 58.60   & 0.42  \\
      & r$^2$SCAN           & 28.845 & 58.034  & 0.26  \\
      & r$^2$SCAN@PGS       & 30.770 & 49.508  & 0.18  \\
      & r$^2$SCAN@LKT       & 29.288 & 55.735  & 0.005 \\
      & MS2                 & 28.693 & 57.693  & 0.13  \\
      & MS2@PGS             & 28.532 & 69.269  & 0.64  \\
      & MS2@LKT             & 29.541 & 66.449  & 0.44  \\
\hline
InP   & Ref.                & 25.231 & 72.00   & 1.42  \\
      & r$^2$SCAN           & 26.290 & 68.757  & 1.18  \\
      & r$^2$SCAN@PGS       & 27.765 & 59.636  & 1.24  \\
      & r$^2$SCAN@LKT       & 26.673 & 65.841  & 0.98  \\
      & MS2                 & 26.289 & 69.269  & 1.15  \\
      & MS2@PGS             & 26.492 & 65.588  & 1.66  \\
      & MS2@LKT             & 26.541 & 66.449  & 1.30  \\
\hline
InSb  & Ref.                & 33.996 & 46.10   & 0.24  \\
      & r$^2$SCAN           & 35.213 & 48.744  & 0.26  \\
      & r$^2$SCAN@PGS       & 38.112 & 36.791  & 0.15  \\
      & r$^2$SCAN@LKT       & 35.889 & 42.296  & 0.015 \\
      & MS2                 & 35.834 & 43.934  & 0.10  \\
      & MS2@PGS             & 34.062 & 47.232  & 0.51  \\
      & MS2@LKT             & 36.038 & 41.132  & 0.47  \\
\end{longtable}
}
\end{*table}


\begin{table}[htbp]
\centering
{\fontsize{10}{12}\selectfont
\caption{\label{tab:coh}
Cohesive energies $E_0$ (in eV/atom) of the LC23 set (the LC20 set extended
with K, Rb, and Cs) computed with r$^2$SCAN and its variants employing the
modified iso-orbital indicators (@PGS, @LKT). Experimental reference values
are taken from Ref.~\cite{OFR2}.}
\begin{ruledtabular}
\begin{tabular}{ccccc}
Material & Ref.(eV /atom) & r$^2$SCAN & r$^2$SCAN@LKT & r$^2$SCAN@PGS \\
\hline
Li    & 1.67 & 1.677 & 1.732 & 1.708 \\
Na    & 1.12 & 1.075 & 1.130 & 1.121 \\
K     & 0.94 & 0.846 & 0.904 & 0.910 \\
Rb    & 0.86 & 0.747 & 0.805 & 0.810 \\
Cs    & 0.81 & 0.681 & 0.763 & 0.754 \\
Ca    & 1.87 & 2.059 & 1.974 & 1.919 \\
Sr    & 1.73 & 1.785 & 1.623 & 1.584 \\
Ba    & 1.91 & 1.991 & 2.000 & 1.830 \\
Al    & 3.43 & 3.597 & 3.303 & 6.957 \\
Cu    & 3.51 & 3.854 & 3.436 & 3.083 \\
Rh    & 5.78 & 5.493 & 5.732 & 4.419 \\
Pd    & 3.93 & 4.162 & 3.880 & 3.203 \\
Ag    & 2.96 & 2.882 & 2.717 & 2.151 \\
C     & 7.55 & 7.890 & 8.437 & 7.474 \\
SiC   & 6.48 & 6.766 & 6.936 & 6.353 \\
Si    & 4.68 & 4.919 & 4.882 & 4.581 \\
Ge    & 3.89 & 3.878 & 4.199 & 3.914 \\
GaAs  & 3.34 & 3.317 & 10.544 & 5.536 \\
LiF   & 4.46 & 4.451 & 4.741 & 4.490 \\
LiCl  & 3.59 & 3.529 & 3.648 & 3.505 \\
NaF   & 3.97 & 4.010 & 4.238 & 4.018 \\
NaCl  & 3.34 & 3.279 & 3.344 & 3.255 \\
MgO   & 5.20 & 5.253 & 5.555 & 6.522 \\
\end{tabular}
\end{ruledtabular}
}
\end{table}

\bibliography{reference}
\bibliographystyle{apsrev}